\documentclass[conference]{IEEEtran}
\IEEEoverridecommandlockouts
\usepackage{cite}
\usepackage{amsmath,amssymb,amsfonts}
\usepackage{algorithmic}
\usepackage{graphicx}
\usepackage{textcomp}
\usepackage{xcolor}
\usepackage{array}
\usepackage{balance}
\usepackage{booktabs}
\usepackage{circuitikz}
\usepackage{cite}
\usepackage{color}
\usepackage{enumitem}
\usepackage{float}
\usepackage{hyperref}
\usepackage{lipsum}
\usepackage{siunitx}
\usepackage{subcaption}
\usepackage[inkscapelatex=false]{svg}
\usepackage{tabularx}
\usepackage{tikz}

\def\BibTeX{{\rm B\kern-.05em{\sc i\kern-.025em b}\kern-.08em
    T\kern-.1667em\lower.7ex\hbox{E}\kern-.125emX}}
\begin{document}

\title{Hardware Implementation of Ring Oscillator Networks Coupled by BEOL Integrated ReRAM for Associative Memory Tasks
\thanks{{This work is funded by EU within the PHASTRAC (grantID: 101092096), by SNSF within ALMOND (grantID: 198612), and by ERC within THERMODON (grantID: 101125031) project. The authors also acknowledge the Binnig and Rohrer Nanotechnology Center (BRNC) at IBM Research Europe.}}
}

\author{
\IEEEauthorblockN{Wooseok Choi\textsuperscript{1,+}, 
Thomas van Bodegraven\textsuperscript{2,+},
Jelle Verest\textsuperscript{2},
Olivier Maher\textsuperscript{1},
Donato F. Falcone\textsuperscript{1}, 
Antonio La Porta\textsuperscript{1},\\
Daniel Jubin\textsuperscript{1},
Bert J. Offrein\textsuperscript{1},
Siegfried Karg\textsuperscript{1},
Valeria Bragaglia\textsuperscript{1,2*},
Aida Todri-Sanial\textsuperscript{2*},
}
\IEEEauthorblockA{\textsuperscript{1}\textit{IBM Research Europe},
Rüschlikon, Switzerland,
\textsuperscript{2}\textit{Eindhoven University of Technology},
Eindhoven, The Netherlands \\
\textsuperscript{+}\textit{Equal contribution}, *Email: vbr@zurich.ibm.com, a.todri.sanial@tue.nl}
}

% This work is funded by EU within the PHASTRAC (grantID: 101092096), by SNSF within ALMOND (grantID: 198612), and by ERC within THERMODON (grantID: 101125031) project
\maketitle

\captionsetup{font=small}

\begin{abstract}
We demonstrate the first hardware implementation of an oscillatory neural network (ONN) utilizing resistive memory (ReRAM) for coupling elements. A ReRAM crossbar array chip, integrated into the Back End of Line (BEOL) of CMOS technology, is leveraged to establish dense coupling elements between oscillator neurons, allowing phase-encoded analog information to be processed in-memory. We also realize an ONN architecture design with the coupling ReRAM array. To validate the architecture experimentally, we present a conductive metal oxide (CMO)/HfO\textsubscript{x} ReRAM array chip integrated with a 2-by-2 ring oscillator-based network. The system successfully retrieves patterns through correct binary phase locking. This proof of concept underscores the potential of ReRAM technology for large-scale, integrated ONNs.
\end{abstract}


\section{Introduction}
\begin{figure}
      \includegraphics[width=\linewidth]{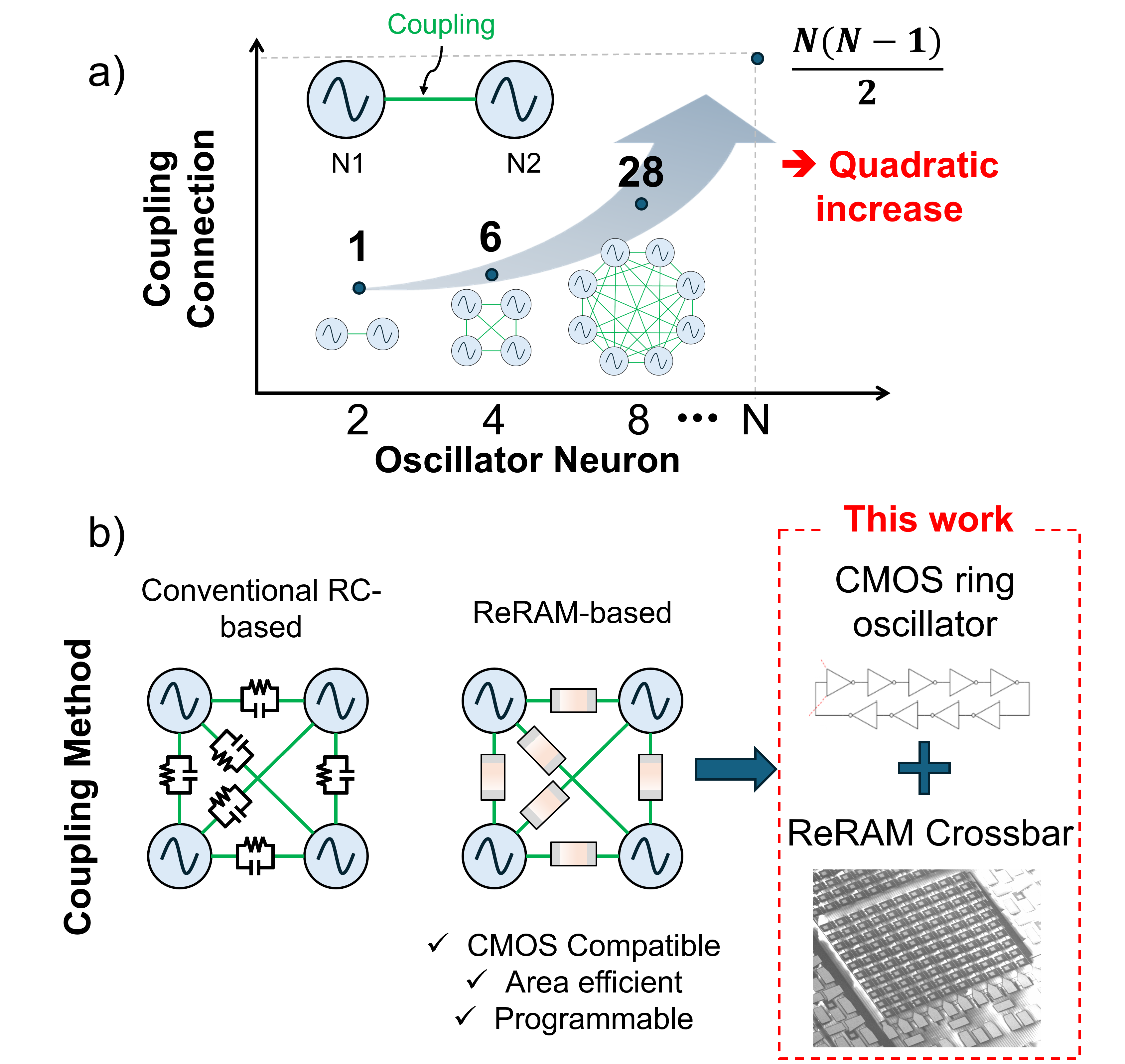}
      \caption{
      (a) Quadratic increase in coupling connections with network size growth. 
      (b) Two coupling methods: conventional passive components vs. ReRAM devices.}
      \label{fig:F1}
    \end{figure}

Brain-inspired oscillatory neural networks (ONNs) are promising neuromorphic computing systems where coupled oscillator neurons compute the phase-encoded analog information simultaneously \cite{maher2024cmos,markus2023, kim2023experimental, kim2023beol}. The coupling devices can be implemented using passive resistors and/or capacitors. However, the number of coupling connections quadratically increases with the growth of network size (\textbf{Fig. \ref{fig:F1}a}). Moreover, the coupling elements must be programmable in resistance and/or capacitance for enabling online learning capability of the ONN hardware \cite{abernot2023oscillatory, todri2024computing, Delacour2024}.
% the advanced silicon technology for low power integrated circuits (ICs) imposes a significant challenge in the area scalability of those passive components \cite{8050350}. 

In this regard, a CMOS compatible, emerging resistive memory (ReRAM) is a key element in realizing the dense coupling elements of ONNs. The ReRAM exhibits promising features suitable for coupling elements, such as low-power operation and multilevel resolution in a scaled area (4F$^2$ footprint) \cite{wooseok, park2023holistic, galetta2024compact, stecconi2024analog}. Especially, in a crossbar array architecture, dense coupling network can be effectively established within a small region. 

This work presents a first proof of concept for ONN hardware using CMOS ring oscillators coupled via a ReRAM crossbar array (\textbf{Fig. \ref{fig:F1}b}), advancing towards an integrated, large-scale ONN system on a chip implementation. We begin by introducing a conductive-metal-oxide (CMO)/HfO\textsubscript{x} ReRAM technology and its integration into a 1 transistor-1 ReRAM (1T1R) array within a 350 nm process node. An architecture for implementing a 2-by-2 ONN on a ReRAM array is also proposed. Through hardware demonstration, we validate that the ring oscillator network can successfully perform associative memory tasks, such as pattern retrieval, by recalling stored patterns using the coupled ReRAM crossbar array.

\section{Experimental}
\begin{figure}
      \includegraphics[width=\linewidth]{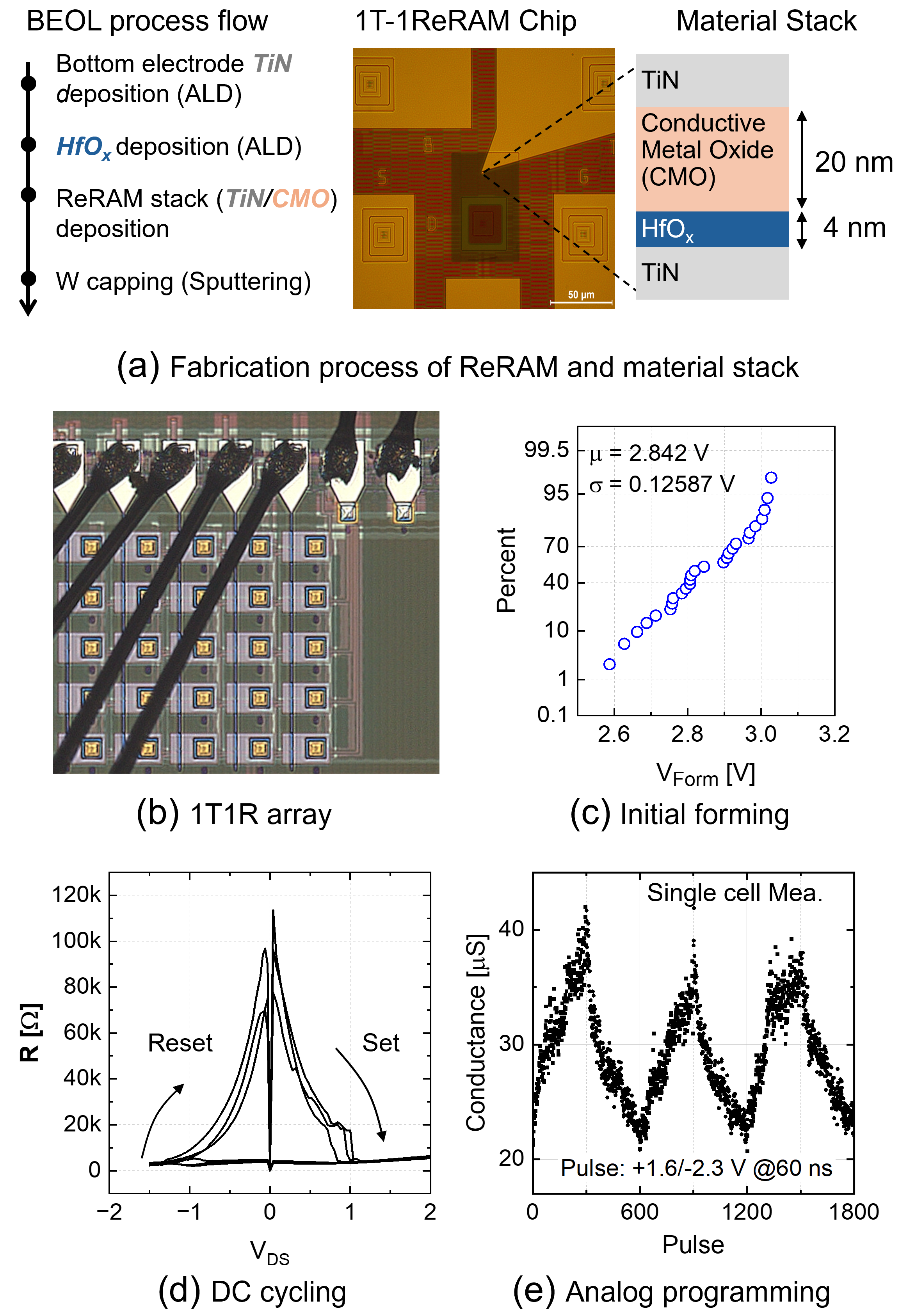}
      \caption{(a) The fabrication of CMO/HfO$_{\rm x}$ ReRAM chip, including (b) 1T1R arrays. (c) The distribution of initial forming voltages and (d) the following resistive switching of ReRAM devices on the wire-bonded $5\times5$ array. (e) The analog programming capability with 60 ns fast pulses is also demonstrated.}
      \label{fig:F2}
    \end{figure}

The fabrication process flow of ReRAM is shown in \textbf{Fig. \ref{fig:F2}a}. For BEOL integration process, TiN and sub-stoichiometric HfO$_{\rm x}$ were grown by ALD at 300 $^\circ$C. Following the deposition of the resistive switching CMO layer, both TiN  and W were subsequently sputtered onto the structure. The fabricated device dimension ranged from 200 nm to 2 \textmu m and devices with a 2 \textmu m dimension were used in this study.
The wire-bonded, $5\times5$ 1T1R crossbar is shown in \textbf{Fig. \ref{fig:F2}b}. The array electrical measurements were conducted using a custom-built setup connected to a National Instrument (NI) system. Nanosecond-fast pulse measurements were conducted for a single cell using a Keithley 4200A machine and pulse measurement units. The assembled ONN hardware was controlled by a custom-designed setup based on the NI system with multichannel analog I/O PXIe cards.

\section{Results}

\subsection{Conductive-Metal-Oxide (CMO)/HfO\texorpdfstring{\textsubscript{x}}{} ReRAM} \label{ReRAM}

\textbf{Fig. \ref{fig:F2}c} shows a distribution of effective forming voltages (V$_{\rm Form}$) of 25 ReRAM devices in the array, where the mean and standard deviation values are 2.84 V and 0.13 V, respectively. The representative DC R-V curves obtained from the wire-bonded array are shown in \textbf{Fig. \ref{fig:F2}d}, with reversible and gradual resistance changes. The resistive switching phenomenon of the device can be attributed to the redistribution of oxygen ions/vacancies in the CMO layer, as described in more detail in our previous studies \cite{falcone2024analytical}. In \textbf{Fig. \ref{fig:F2}e}, we present the analog programming capability of CMO/HfO$_{\rm x}$ ReRAM by applying 60 ns fast voltage pulses with identical amplitudes +1.6 V for partial set and -2.3 V for partial reset, respectively. The number of programming pulses in a cycle was 300. These results demonstrate its potential for online learning applications in neural network hardware. 
It is worthy to note that there are a few important considerations for the coupling ReRAM devices in the ONN that need to be addressed during architecture design. 
    \begin{itemize}
        \item The resistive states of ReRAM can be changed during computing, when a large voltage difference between two oscillators is applied to the coupling ReRAM.
        \item The non-linear dependence of the ReRAM resistance on the applied voltage can incur computing errors, since the oscillating voltages are dynamically applied to the ReRAM devices.
    \end{itemize}
    
\subsection{2-by-2 ONN Design with ReRAM Array Coupling}

    \begin{figure} [!h]
      \includegraphics[width=\linewidth]{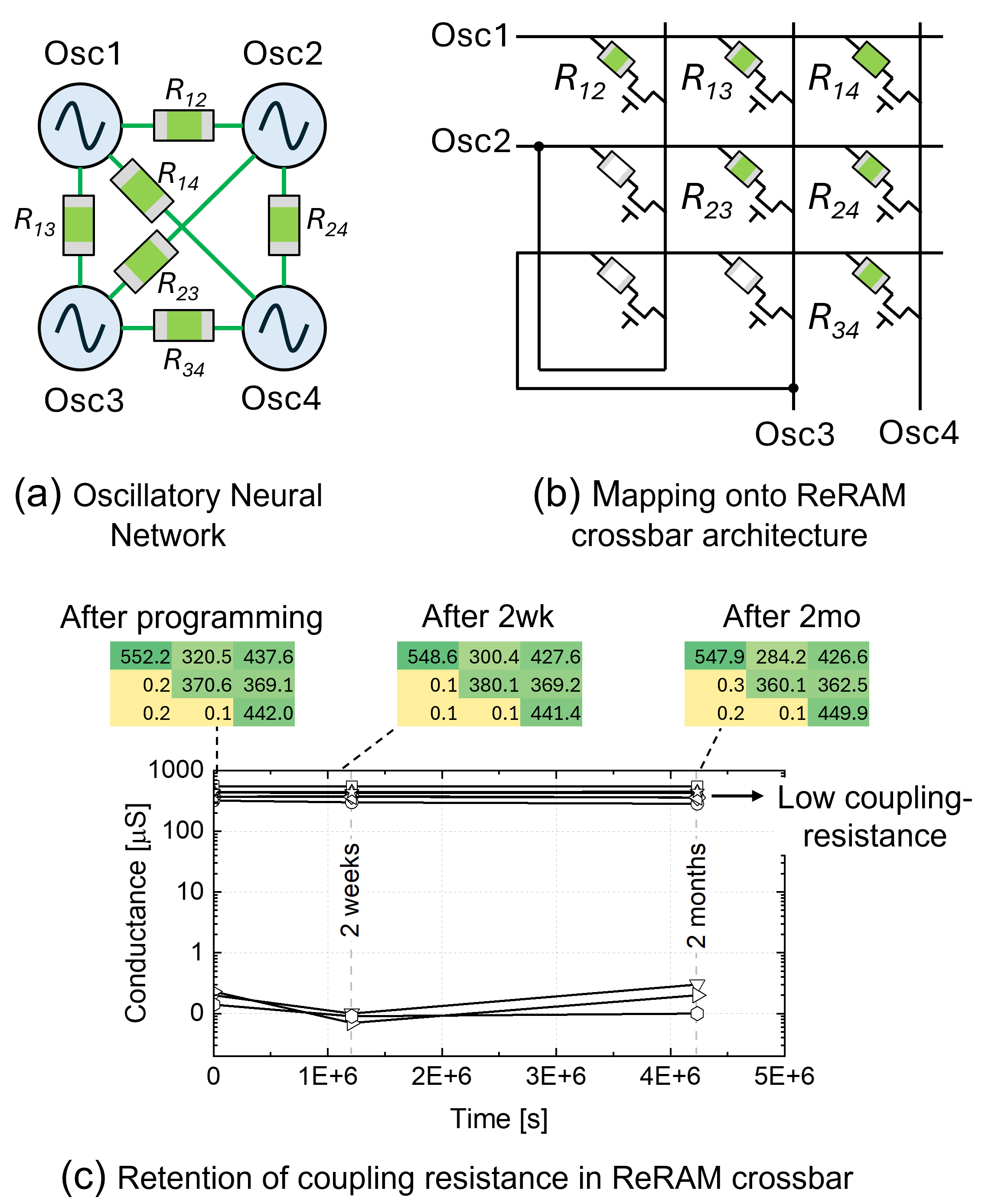}
      \caption{(a) 2-by2 ONN and (b) its architecture implementation with ReRAM crossbar array. The experimental retention data of the implemented coupling resistance are shown in (c).}
      \label{fig:F3}
    \end{figure}
    
As shown in \textbf{Fig. \ref{fig:F3}a}, a 2-by-2 ONN inspired by Hopfield neural networks was designed to implement associative memory for pattern retrieval of 4-pixel image, which is further discussed in the following sections. Six connections are needed for coupling of a 2-by-2 ONN. The ONN architecture using ReRAM crossbar array is shown in \textbf{Fig. \ref{fig:F3}b}. When mapping the 2-by-2 ONN to the ReRAM crossbar, neuron 2 and 3 have to be connected to two array inputs such that each oscillator is coupled with all other oscillators, as shown in \textbf{Fig. \ref{fig:F3}a}. The coupling devices (green) are programmed to a low resistance of approximately 2 k$\Omega$, while redundant devices (white) did not undergo the
initial forming process and hence have a high resistance of approximately 10 M$\Omega$. 
% To allow for a strong coupling, the lowest resistance was chosen for the coupling devices, while the highest resistance was chosen for the redundant devices, creating a weak coupling. 
\textbf{Fig. \ref{fig:F3}c} shows experimental results of the programmed array, demonstrating robust data retention over two months. 

\subsection{Ring Oscillator Neurons}

    \begin{figure}
      \includegraphics[width=\linewidth]{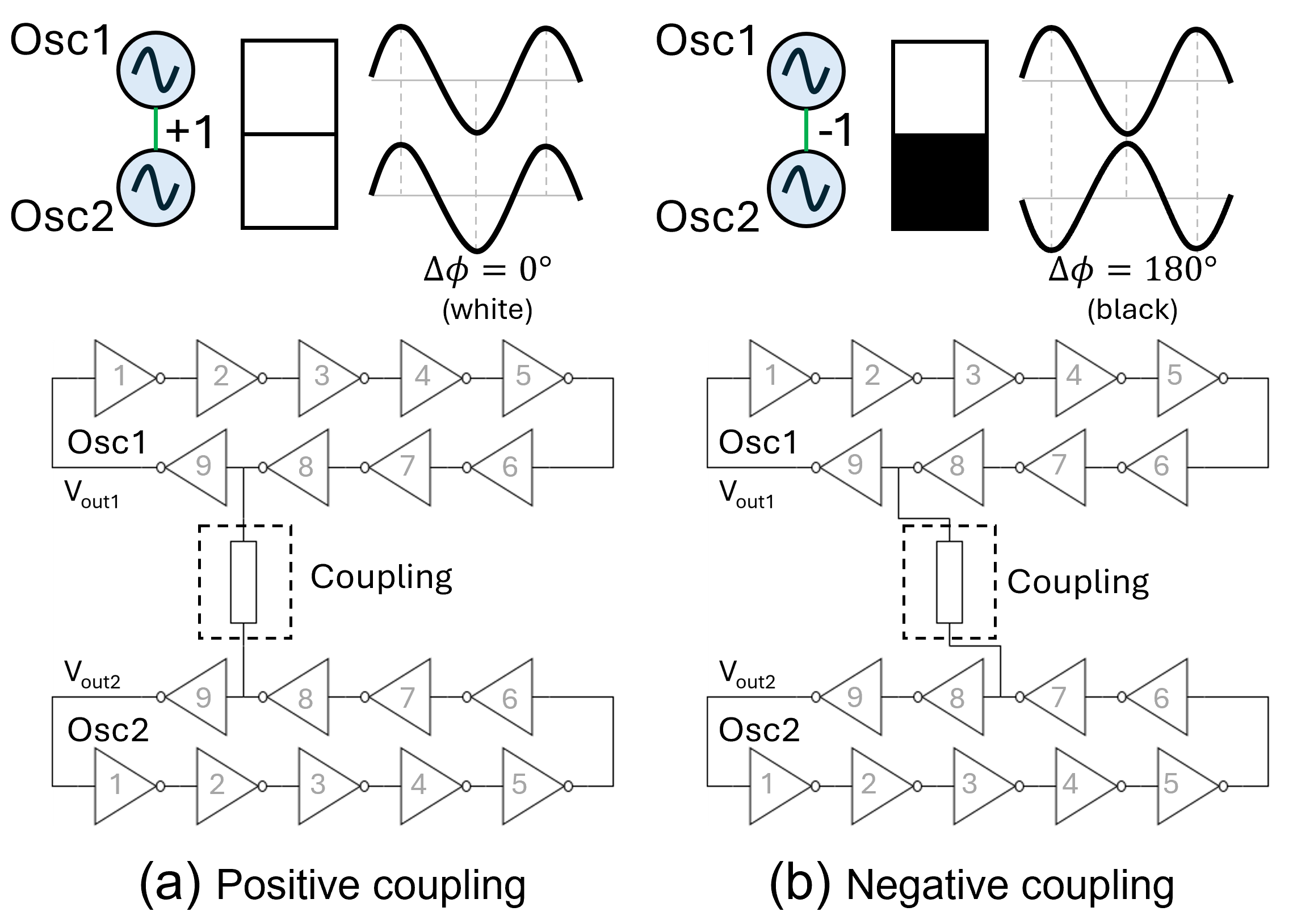}
      \caption{Examples of resulting phase locking depending on (a) positive and (b) negative coupling between two CMOS ring oscillators (9-stage).
      }
      \label{fig:F4}
    \end{figure}
    
After forming the desired connections in the array (\textbf{Fig. \ref{fig:F3}}), the polarity of coupling weights between neurons were defined. When the first oscillator neuron (Osc1) is considered as a reference (\textbf{Figs. \ref{fig:F4}a and b}), phase locking of the other neuron result in either in-phase ($\Delta\phi=0^\circ$) or out-of-phase ($\Delta\phi=180^\circ$), depending on the coupling direction. With 9-stage CMOS ring oscillators, the symmetric (or asymmetric) coupling configuration between the two neurons represents positive (or negative) coupling, causing the in-phase (or out-of-phase) relationship. For the coupling, we used the seventh and eighth stages behind the output inverter in each neuron  (\textbf{Fig. \ref{fig:F4}}).
Note that a 47 k$\Omega$ series resistor is added to each output of the coupled inverter to alter the output impedance, enabling frequency locking with the coupled ReRAM resistance.
Additionally, the resistors act as a voltage divider with the coupling ReRAM. This helps avoid issues, i.e., state disturbance of ReRAM devices due to voltage stresses and computation errors due to non-linear device resistance (see \autoref{ReRAM}).

% For a 2-by-2 pixel image, there are six meaningful patterns: horizontal pattern (HP), vertical pattern (VP), diagonal pattern (DP) and its inverses (HPI, VPI, DPI). \autoref{tab:pattern_config} shows the connection configuration of 1T1R crossbar array inputs to realize certain patterns. $RO_{11}$ means ring oscillator 1 to be connected after the first stage, $RO_{12}$ means ring oscillator 1 to be connected after the second stage. For the 3-by-3 crossbar array, $R_{BD}$, $R_{CD}$ and $R_{CE}$ were redundant, and therefore uninitialized ($\sim$11-15 M$\Omega$).

%     \begin{table}
%     \centering
%     \caption{Configurations for the patterns}
%     \label{tab:pattern_config}
%     \begin{tabular}{c c c c c c c}
%     \toprule
%     Voltage node & HP & VP & DP & HPI& VPI & DPI\\ 
%     \midrule
%     $V_A$ & $RO_{11}$ & $RO_{11}$ & $RO_{11}$ & $RO_{12}$ & $RO_{12}$ & $RO_{12}$ \\ 
%     $V_B$ & $RO_{21}$ & $RO_{22}$ & $RO_{22}$ & $RO_{22}$ & $RO_{21}$ & $RO_{21}$ \\
%     $V_C$ & $RO_{32}$ & $RO_{31}$ & $RO_{32}$ & $RO_{31}$ & $RO_{32}$ &  $RO_{31}$\\
%     $V_D$ & $RO_{21}$ & $RO_{22}$ & $RO_{22}$ & $RO_{22}$ & $RO_{21}$ &  $RO_{21}$\\
%     $V_E$ & $RO_{32}$ & $RO_{31}$ & $RO_{32}$ & $RO_{31}$ & $RO_{32}$ &  $RO_{31}$\\
%     $V_F$ & $RO_{42}$ & $RO_{42}$ & $RO_{41}$ & $RO_{41}$ & $RO_{41}$&  $RO_{42}$\\
%     \bottomrule
%     \end{tabular}
%     \end{table}
    
\subsection{ONN Hardware Demonstration}

    \begin{figure} [t]
        \includegraphics[width=\linewidth]{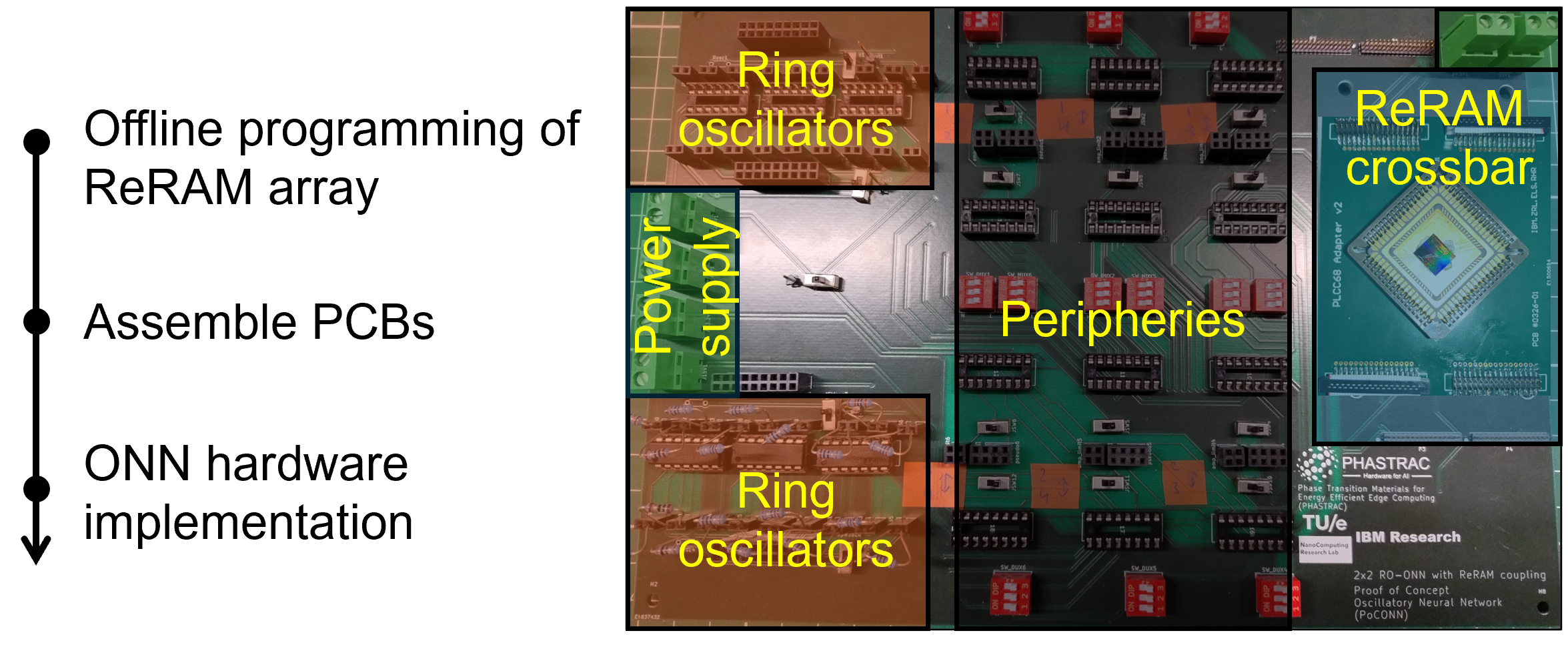}
        \caption{Operation flow of the experiments and assembled ONN hardware system.}
        \label{fig:F5}
    \end{figure}

    \begin{figure} [t]
        \includegraphics[width=\linewidth]{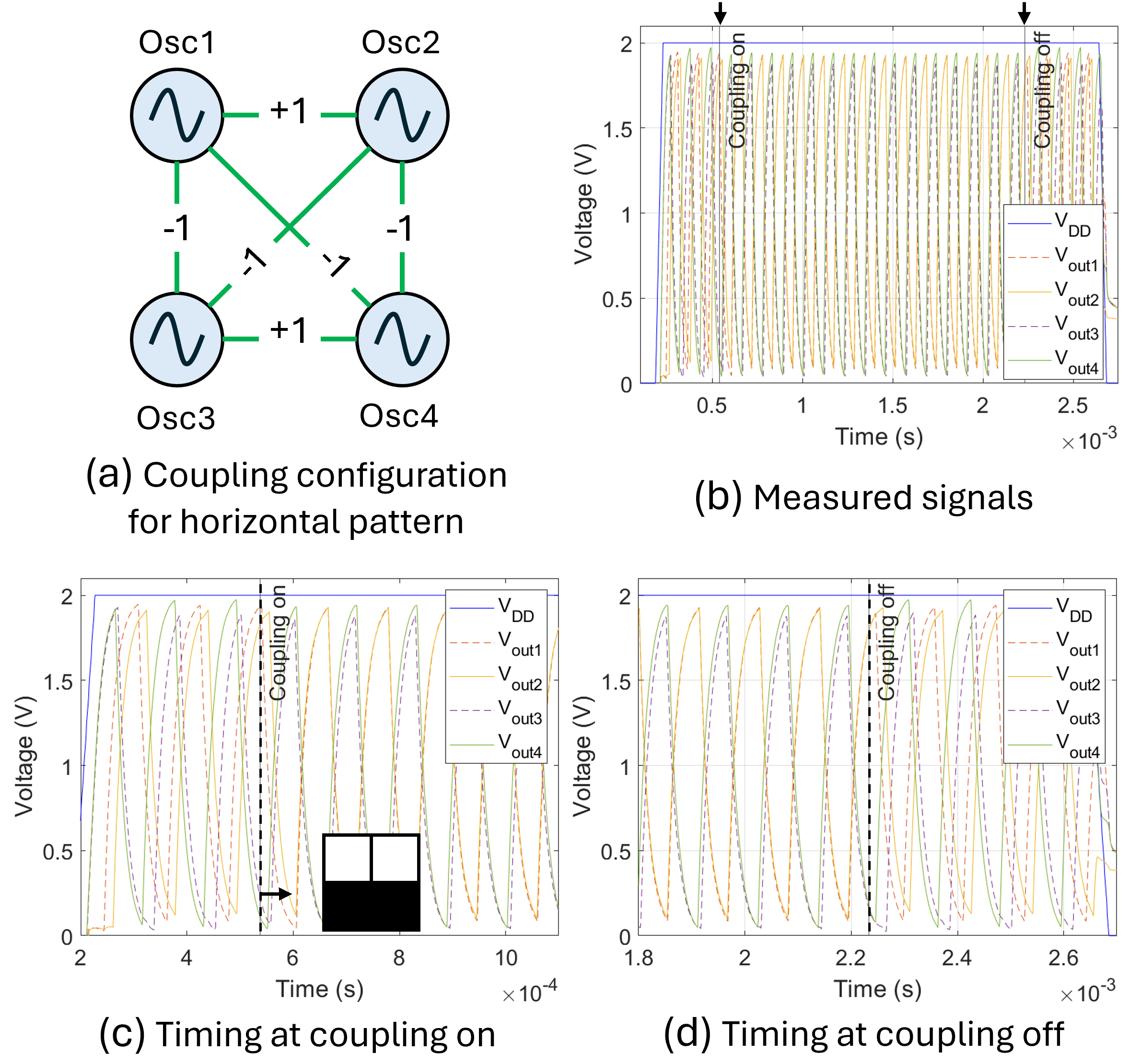}
        \caption{ Experimental results for horizontal pattern retrieval of the ONN coupled by a ReRAM array chip.}
        \label{fig:F6}
    \end{figure}

A printed circuit board (PCB) was designed according to the verified ONN circuit architecture through Cadence Virtuoso simulations (\textbf{Fig. \ref{fig:F5}}). After the offline programming of the ReRAM array, the hardware was assembled. The left section is dedicated to four ring oscillators with series resistors. The middle section with peripheries is where the coupling direction between oscillators can be set (\textbf{Fig. \ref{fig:F4}}). The right section shows the space for mounting the ReRAM crossbar array chip on the PCB. 
During ONN inference for the pattern retrieval, the frequencies of each oscillator are first synchronized. Then, the phase of each oscillator is compared to that of a chosen reference oscillator, which represents a white pixel in the experiments. Oscillators with an in-phase relation indicate a white pixel ($\Delta\phi=0^\circ$), while oscillators with an out-of-phase relation indicate a black pixel ($\Delta\phi=180^\circ$).  

    \begin{figure}
      \includegraphics[width=\linewidth]{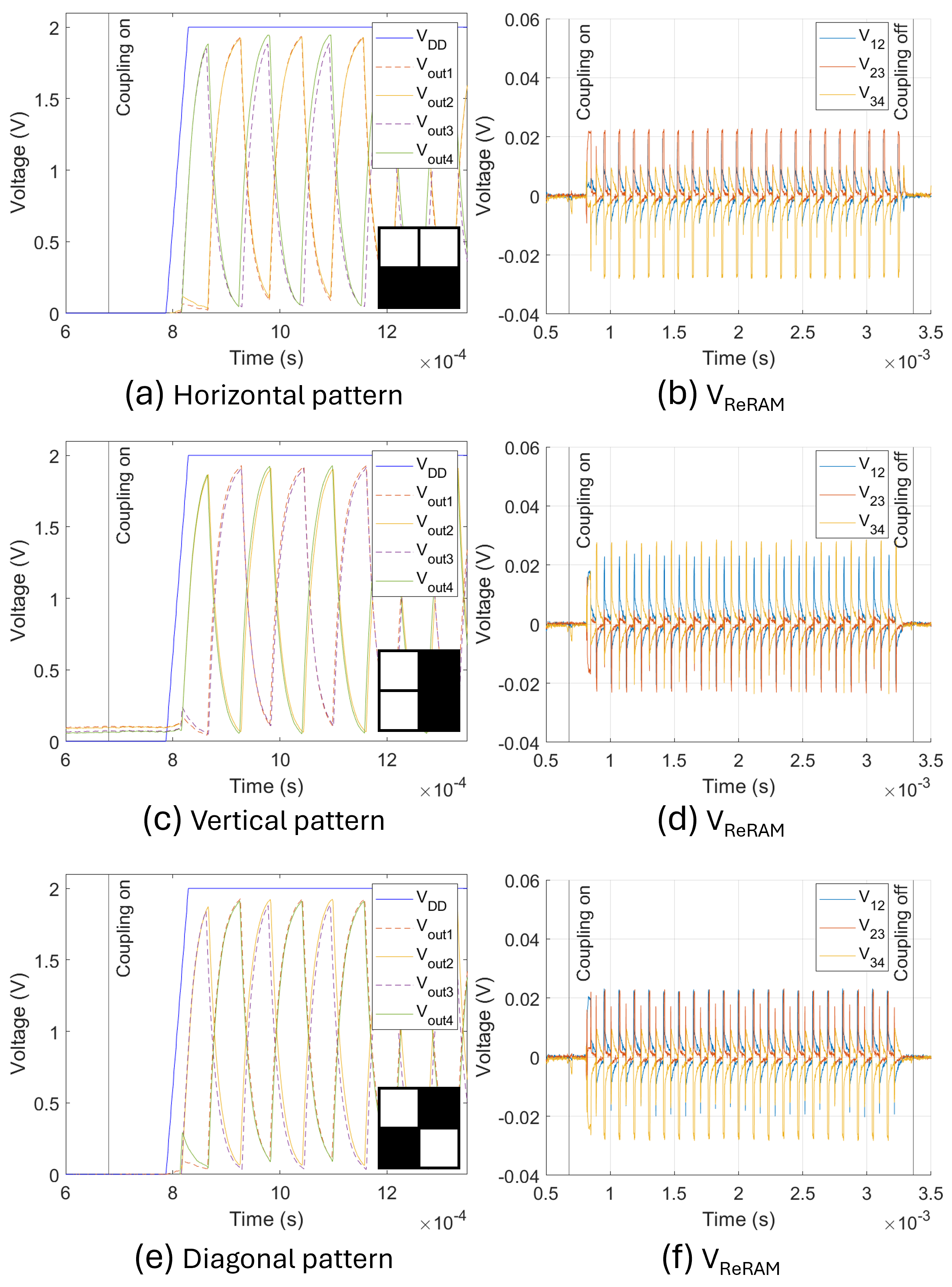}
      \caption{Experimental results of pattern retrieval with the ONN hardware for horizontal, vertical, and diagonal patterns.}
      \label{fig:F7}
    \end{figure}  

\textbf{Fig. \ref{fig:F6}a} shows an example of the coupling configuration stored for horizontal pattern retrieval. \textbf{Figs. \ref{fig:F6}b} shows the experimental results, where the ReRAM coupling was dynamically switched on and off, while the neurons were freely oscillating with different frequencies. The switch was activated by applying 4.5 V to the gate of connected transistors in 1T1R array (coupling on). \textbf{Fig. \ref{fig:F6}c and d} show the results around the timing of each switch operation. We can observe that each neuron began oscillating again at its own intrinsic frequency after disconnecting the coupling (\textbf{Fig. \ref{fig:F6}d}). Based on the top-left reference neuron (Osc1), the top-right neuron (Osc2) is oscillating in-phase ($\Delta\phi_{12}=0^\circ$), while the others (Osc3 and Osc4) are oscillating out-of-phase ($\Delta\phi_{13}=\Delta\phi_{14}=180^\circ$). The operating frequency during synchronization was 8.6 kHz. 

The experiments for horizontal, vertical, and diagonal patterns were also shown  in \textbf{Fig. \ref{fig:F7}}. We observe that the pattern retrieval efficiently happens within one oscillation period for all three experiments (\textbf{Fig. \ref{fig:F7}a, c, and e}). The synchronized waveforms across the operation time indicate that there was no resistance disturbance in coupling ReRAM devices during operation. In \textbf{Fig. \ref{fig:F7}b, d, and f}, the voltage drop across the coupling ReRAM devices connected to the reference neuron is also shown for each pattern. Although the ReRAM devices in this study were pre-programmed for implementing the specific coupling connections, online learning capability of ONN will be further explored by leveraging the programmable features of the analog ReRAM (Fig. \ref{fig:F2}e).

\section{Conclusion}
We demonstrated the first hardware implementation of an ONN coupled by a ReRAM crossbar array. The BEOL-integrated 1T1R arrays were introduced for enabling dense coupling networks in ONNs. We also described an architecture for implementing ONNs onto a crossbar array. By assembling the ReRAM chip with oscillator neurons, we successfully demonstrated pattern retrieval of a 2×2 image. The results represent a significant step toward the development of integrated ONN architectures, leveraging emerging ReRAM technology. This work opens up a viable design path for scaling up ONN implementation and enabling both inference and dynamic programming for online and on-chip learning.

% \section*{Acknowledgment}
% {The authors acknowledge the Binnig and Rohrer Nanotechnology Center (BRNC) at IBM Research Europe - Zurich.}

\balance
\bibliographystyle{IEEEtran}
\bibliography{INT}
\end{document}